# Humans and Technology for Inclusive Privacy and Security

**Organizers & Moderators:**
Sanchari Das, University of Denver
Robert S. Gutzwiller, Arizona State University
Rod D. Roscoe, Arizona State University

**Panelists:**
Prashanth Rajivan, University of Washington
Yang Wang, University of Illinois, Urbana-Champaign
L. Jean Camp, Indiana University Bloomington
Roberto Hoyle, Oberlin College

Computer security and user privacy are critical issues and concerns in the digital era due to both increasing users and threats to their data. Separate issues arise between generic cybersecurity guidance (i.e., protect all user data from malicious threats) and the individualistic approach of privacy (i.e., specific to users and dependent on user needs and risk perceptions). Research has shown that several security- and privacy-focused vulnerabilities are technological (e.g., software bugs (Streiff, Kenny, Das, Leeth, & Camp, 2018), insecure authentication (Das, Wang, Tingle, & Camp, 2019)), or behavioral (e.g., sharing passwords (Das, Dingman, & Camp, 2018); and compliance (Das, Dev, & Srinivasan, 2018) (Dev, Das, Rashidi, & Camp, 2019)). This panel proposal addresses a third category of sociotechnical vulnerabilities that can and sometimes do arise from non-inclusive design of security and privacy. In this panel, we will address users' needs and desires for privacy. The panel will engage in in-depth discussions about value-sensitive design while focusing on potentially vulnerable populations, such as older adults, teens, persons with disabilities, and others who are not typically emphasized in general security and privacy concerns. Human factors have a stake in and ability to facilitate improvements in these areas.

## INTRODUCTION

The newly-formed Cybersecurity Technical Group of the Human Factors and Ergonomics Society (HFES) recognizes the essential importance of human factors and ergonomics in cybersecurity research, practice, and application. Issues of computer security and privacy are fundamentally related to peoples' needs, goals, beliefs, and interactions. This group has previously examined sponsor perspectives (Gutzwiller, Cosley, Ferguson-Walter, Fraze, & Rahmer, 2019) and human-machine teaming (Paul, Blaha, Fallon, Gonzalez, & Gutzwiller, 2019), but the intersection of cybersecurity and human concerns is vast (Gutzwiller, Fugate, Sawyer, & Hancock, 2015).

One example set of concerns pertains to inclusion, equity, and social justice--themes that have gained increasing prominence in the work of HFES. The HFES Diversity and Inclusion Committee strives for meaningful representation and participation of diverse individuals in HFES (Chiou, Wooldridge, Price, Mosqueda, & Roscoe, 2017) and has advocated strongly for addressing these concerns as outcomes of our work (Roscoe, Chiou, & Wooldridge, 2019; Wooldridge et al., 2019, 2018). Similarly, the recently-formed HFES Societal Impact Committee is charged with exploring how HFES–including individual members, technical groups, and the organization as a whole–can contribute to *beneficial* outcomes and advancement at a societal level (e.g., public and cultural awareness, economic and public policy, legal trends, and consumer products).

In response to these dual charges, this panel will discuss ***how to design computing security and privacy in ways that are usable and applicable to all people, including vulnerable and marginalized populations***. For example, persons with limited vision or manual dexterity are excluded from security measures that rely on visual images or passwords typed on tiny virtual keyboards with no haptic feedback. Similarly, older adults and LGBTQIA+ individuals may be targeted specifically or differently by cyberattacks and may experience more severe consequences. Cybersecurity methods and training that ignore these aspects of human identity and experience will necessarily be incomplete. However, developing robustly inclusive tools will not only help vulnerable populations, but also benefit all users.

Panelists will discuss how methods, expertise, and research can advance cybersecurity and privacy technology while meeting the needs of diverse users and beliefs. We bring together security and privacy experts from both academia and industry who approach human factors from a mix of backgrounds, experience, and research traditions. The panelists will highlight current work, offer guidance for applying human factors and ergonomics expertise to these issues, and discuss avenues for dissemination.

In the following sections, the panelists provide statements that briefly describe the concepts and viewpoints they will address in the panel.

**Inclusive Security and Privacy Education and Training**

*Prashanth Rajivan, PhD*
*Industrial and Systems Engineering*
*University of Washington*

Unlike *physical* security and privacy, which people have been dealing with for several thousands of years, information security and privacy continue to be abstract and ambiguous concepts for most individuals. Even computer savvy, college-educated individuals demonstrate difficulty in differentiating between important, everyday security concepts, such as differences between a spam email (Das, Abbott, Gopavaram, Blythe, & Camp, 2020) and a phishing attack (Das, Kim, Tingle, & Nippert-Eng, 2019). People regularly place inappropriate optimism on accessible security controls (e.g., security certificates or antivirus software) while discounting other essential and proactive security controls (e.g., software updates (Stallings, Brown, Bauer, & Bhattacharjee, 2012)). This neglect leads to detrimental security and privacy behaviors (Rajivan, Moriano, Kelley, & Camp, 2017).

Researchers and practitioners are continuing efforts to increase education and raise awareness among the general public about the risks of information security and privacy (Kirlappos, Sasse, & Harvey, 2012). Such efforts include public service announcements, news messages about the latest security attacks, online education materials, on-the-job security awareness training, or formal education in security and privacy. However, existing security training and education efforts are not available and/or do not cater to individuals with different abilities, characteristics, needs, identities, and values. The challenges in such cases are likely more pronounced than the general public.

Inclusiveness should be expected and demanded in all forms of security awareness and training procedures. However, there are two main challenges to developing inclusive security education and training: (1) tailoring security training to suit individual population needs and abilities, and (2) making training available and accessible to everyone, not just to individuals who are financially and socially advantaged. Technology in the form of human-aware, human-centered machine learning is amenable to developing personalized security training, but the challenge lies in understanding the challenges faced by different populations.

As one example, consider the broad communities of people with autism, learning disabilities, and/or Attention Deficit Hyperactivity Disorder (ADHD). Some individuals with autism have difficulty understanding abstract concepts, ambiguity, and nuances of social communication -- all of which are relevant to deception and phishing attacks. Training these individuals to detect phishing emails or spoofed security certificates using existing methods is likely to fail. Research has shown that too many colors or sounds can cause visual and auditory overstimulation, which in turn significantly affects learning. Thus, care is necessary for designing the appropriate interface. Similar challenges and creative solutions are needed for individuals with various learning or attention challenges.

**Design for Inclusive Security and Privacy**

*Yang Wang, PhD*
*Information Science*
*University of Illinois at Urbana-Champaign*

Existing end-user privacy mechanisms are often designed without considering the diverse user populations beyond the convenient sample space. There have been few studies that explore the technological perceptions of users belonging to different cultural, technical, and physical statuses. Thus, exploring other populations is very critical. One such study by Briggs and Thomas (2015) identified common and different factors when people think about future technologies.

However, we notice a critical literature gap in the security and privacy perceptions of underserved populations. In the privacy and security research domain, underserved populations may include persons with disabilities, children, older adults, and people from non-Western developing countries. As a result, we often find non-inclusive designs in the privacy and security domain due to many biased assumptions. These inappropriate assumptions could lead to significant challenges for underserved users to utilize privacy mechanisms. Through our research, we seek to design tools and technologies that empower users to protect their privacy. However, misaligned designs often make it problematic for users to use such tools, techniques, and mechanisms. Thus, difficulties in effectively using these mechanisms could, in turn, make the underserved users more vulnerable to various privacy risks.

Along these lines of discussion, the concept of designing inclusive privacy and security becomes incredibly critical. Inclusive privacy and security design aims to address a wide range of users with diverse abilities, characteristics, needs, and values. Such research can build upon previously explored ideas, such as value-sensitive design (VSD) and accessible design. VSD is a standard approach that incorporates and promotes values, such as user autonomy, freedom from bias, privacy, and trust in system framework structure (Friedman, 2008). Insights from the field of accessible computing can also be useful in making security and privacy designs inclusive to a wide range of user populations. For example, Wobbrock and colleagues propose an ability-based model, which shifts the view from focusing on people's disabilities to their abilities (Wobbrock, Kane, Gajos, Harada, & Froehlich, 2011).
Inclusiveness is desirable and valuable. Security and privacy designers need to state their value judgments and justify their design decisions, especially when there are conflicts (e.g., national security vs. personal privacy) (Wang, 2018).

**Security for Older Adults**

*L. Jean Camp, PhD*
*Informatics*
*Indiana University Bloomington*

A core part of working in security and privacy infrastructure is making sure that everyone's voice is heard. Diversity is a required component for a successful outcome of such research. The majority of research on computer security and data privacy focuses on a convenience sample, often ignoring diverse and vulnerable populations. However, in computer security, if we look at a vulnerable sample of older adults, we find that elders are distinct in terms of their experience with computers, cognitive constraints, and (offline) susceptibility to fraud (Caine et al., 2011; Das, Kim, Jelen, et al., 2019).

To consider the implications, the percentage of the population over sixty-five will be 22% in 2030. Individuals in this population own over one-third of the value of U.S. stock exchanges combined and more than one-tenth of all bonds. As of 2019, there are more people on the planet over 65 than under 5, likely for the first time in human history. People with the design constraints considered in this research are also populations who could arguably obtain the most benefit from MFA when it is designed to their needs. Security policies for the real world should integrate diversity. For example, a standard solution to multiple challenges in the niche security area of authentication and access management is the idea of "correct names" or "real names." Most illustratively, Google's requirement for "actual" names in the launch of G+ was reversed based on the harm of silencing individuals with "real" names and the fact that it was ineffectual in reducing trolling (Cho & Acquisti, 2013). Ironically, Facebook proposed exactly this policy to address foreign interference in American partisan politics. The "real name" failure illustrates the extent that authentication work impinges on the identities of people, as opposed to digital artifacts and organizations. Lack of inclusion has not been simply a failure in diversity but also a failure in science.

**Security for Underrepresented Minorities**

*Roberto Hoyle, PhD*
*Computer Science*
*Oberlin College*

It is a moral imperative to make sure that the benefits of privacy and security research apply to everyone, not just the most educated or well-off among us. Unfortunately, security research in privacy and security has mostly focused on a subset of people and assumed that the findings will apply to all. Such biases have led to inequalities in protections, and in some cases, even exposed individuals to significant risk. This panel is not just a theoretical call for diversity in research. We have to address the practical consequences of such research bias. Facebook, for example, instituted a "real name" policy to discourage fake accounts. This policy directly affects transgender individuals who may be experimenting with a new identity yet are forced to keep their legal "dead" name in full view. Such policies can also have the unintended consequence of outing a transgender individual to people that they would like to be connected with on a social network.

A "one size fits all" approach to privacy and security policies will not work. These endeavors need to be enacted in a way that will not marginalize or discriminate against underrepresented minorities. Fortunately, this issue has been brought to light, and now researchers are investigating how LGBTQIA+, elderly, and other groups feel about privacy protections and how we can design systems to help them better (Lerner, He, Kawakami, Zeamer, & Hoyle, 2020).

## CONCLUSION

In this panel and in previous research, we note that human factors are a critical component in computer security and data privacy. Given the advanced digital age, it is particularly important to study the different aspects of inclusivity and diversity in cybersecurity and privacy studies. Notably, this panel attempts to discuss inclusive privacy and security in general, while also exploring the assumptions and expectations of such research. We will also detail the design challenges and values of this research and address particular cases of vulnerable populations.